\documentclass{aa}
\usepackage{txfonts}

\newcommand{\ce}{\ifmmode {\cal E} \else ${\cal E}$\ \fi}
\newcommand{\kms}{\ifmmode {\rm km\ s}^{-1} \else km s$^{-1}$\ \fi}
\newcommand{\ergs}{\ifmmode {\rm erg\ s}^{-1} \else erg s$^{-1}$\ \fi}
\newcommand{\tes}{\ifmmode \tau_{\rm es} \else $\tau_{\rm es}$\ \fi}
\newcommand{\tk}{\ifmmode \tau_{\rm K} \else $\tau_{\rm K}$\ \fi}
\newcommand{\vfwhm}{\ifmmode V_{\mbox{\tiny FWHM}} \else
            $V_{\mbox{\tiny FWHM}}$\fi}
\newcommand{\msun}{\ifmmode M_{\odot} \else $M_{\odot}$\ \fi}
\newcommand{\afe}{\ifmmode {\mathcal A_{\rm Fe}} \else${\mathcal A_{\rm Fe}}$\ \fi}

\newcommand{\feii}{\rm Fe {\sc II}\ }
\newcommand{\mgii}{\rm Mg {\sc II}}
\newcommand{\civ}{{\rm  C \texttt{IV}}}
\newcommand{\lb}{\ifmmode L_{\rm Bol} \else $L_{\rm Bol}$\ \fi}
\newcommand{\ledd}{\ifmmode L_{\rm Edd} \else $L_{\rm Edd}$\ \fi}
\newcommand{\lx}{\ifmmode L_{\rm 2-10keV} \else  $L_{\rm 2-10keV}$\ \fi}
\newcommand{\hb}{\ifmmode \rm H\beta \else H$\beta$\ \fi}
\newcommand{\ha}{\ifmmode \rm H\alpha \else H$\alpha$\ \fi}
\newcommand{\oiii}{{\sc [O iii]}}

\newcommand{\mbh}{\ifmmode M_{\rm BH}  \else $M_{\rm BH}$\ \fi}
\newcommand{\lv}{\ifmmode \lambda L_{\lambda}(5100\AA) \else $\lambda L_{\lambda}(5100\AA)$\ \fi}
\newcommand{\mdot}{\ifmmode \dot{m} \else \dot{m} \fi }
\newcommand{\llog}{\ifmmode {\rm log} \else {\rm log} \fi }

\usepackage{graphicx}
\begin{document}

\title{Supermassive Black Hole Masses  in Type II Active Galactic Nuclei with
Polarimetric Broad Emission Lines}
   \author{Shi-Yan Zhang
      \inst{1}
   \and Wei-Hao Bian
      \inst{1,2}
   \and Ke-Liang Huang
      \inst{1}
      }

   \institute{Department of Physics and Institute of Theoretical
Physics, Nanjing Normal University, Nanjing 210097, China
        \and
Key Laboratory for Particle Astrophysics, Institute of High Energy
Physics, Chinese Academy of Sciences, Beijing 100039, China
             }
\abstract {} {Type II AGNs with polarimetric broad emission line
provided strong evidence for the orientation-based unified model for
AGNs. We want to investigate whether the polarimetric broad emission
line in type II AGNs can be used to calculate their central
supermassive black hole (SMBH) masses, like that for type I AGNs.}
{We collected 12 type II AGNs with polarimetric broad emission line
width from the literatures, and calculated their central black hole
masses from the polarimetric broad line width and the isotropic
\oiii\ luminosity. We also calculate the mass from stellar velocity
dispersion, $\sigma_*$, with the $\mbh-\sigma_*$ relation.} {We find
that: (1) the black hole masses derived from the polarimetric broad
line width is averagely larger than that from the $\mbh- \sigma_*$
relation by about 0.6 dex, (2) If these type II AGNs follow
$\mbh-\sigma_*$ relation, we find that the random velocity can't not
be omitted and is comparable with the BLRs Keplerian velocity. It is
consistent with the scenery of large outflow from the accretion disk
suggested by Yong et al. } {}

\titlerunning{Supermassive Black Hole Masses in Type II AGNs}
 \maketitle

\section{INTRODUCTION}
The standard paradigm for active galactic nuclei (AGNs) posits an
accretion disk surrounding a central supermassive black hole
(SMBH), along with other components, such as the broad-line
regions (BLRs), narrow-line regions (NLRs), jet, and torus (e.g.,
Antonoucci 1993). The black hole mass ($\mbh$) is an important
parameter for us to understand the nuclear energy mechanics, the
formation and evolution of SMBH and the galaxies (e.g., Rees 1984;
Tremaine et al. 2002).

In the past decade, one of the most progresses in the study of AGNs
is that the masses of SMBHs can be calculated by using the width of
the broad emission lines from BLRs (e.g., $\hb$, $\ha$, $\mgii$,
$\civ$) by the reverberation mapping method and several
corresponding empirical relations (e.g., Kaspi et al.2000, Bian \&
Zhao 2004, Greene \& Ho 2006). In the orientation-based unified
model for AGNs, the distinction between type I AGNs and type II AGNs
depends upon whether the central engine and BLRs are viewed directly
(type I) or are obscured by the circumnuclear torus (type II).
Because of the absence of broad emission lines in the spectrum of
type II AGNs, above methods for the mass calculation are only
applicable to type I AGNs. The SMBH mass in the center of type II
AGNs generally may be estimated by the $\mbh- \sigma_*$ relation
(e.g., Kauffmann et al. 2003; Bian \& Gu 2007).

With spectro-polarimetric observation, some type II AGNs show
hidden BLRs (HBLRs) and some do not (e.g., Antonucci $\&$ Miller
1985; Tran 1995). It is still not clear what kind of physical
process is related to the presence of HBLRs in type II AGNs (e.g.,
Bian \& Gu 2007). In this paper, we calculate the SMBH mass in
type II AGNs with HBLRs by using broad emission lines in their
polarimetric spectrum as well as the $\mbh- \sigma_*$ relation. In
$\S$ 2, we briefly introduce our sample. $\S$ 3 introduces the
methods to calculate the SMBH masses. $\S$ 4 is the data analysis.
Our results and discussions are given in $\S$ 5. The last section
is our conclusion. All of the cosmological calculations in this
paper assume $H_{\rm 0}$=70$\kms$ $Mpc^{-1}$, $\Omega_{M}$=0.3,
$\Omega_{\wedge }$=0.7.

\section{Sample}
We collected a sample of 12 type II AGNs with HBLRs from
literature, which provide us with the hidden broad line width
(Table 1). In this sample, there are 8 type II AGNs with redshifts
z$<$0.05 (Tran 1995), the rest are in the redshift range
0.3$<$z$<$0.8, which are from Zakamska et al. (2007).

For Mrk 3 and Mrk 348, Tran (1995) used a sixth- or seventh-order
polynomial to extract polarimetric broad line profile for their high
s/N. For the remaining objects in Tran (1995), multiple Gaussians
are used to extract polarimetric broad line profile for measuring
the full widths at half-maximum (FWHM). For four objects in Zakamska
et al. (2007), FWHM for polarimetric broad emission lines are
obtained through Gaussian fits. These polarimetric broad emission
lines are obvious because Zakamska et al. (2007) adopted a
sensitivity limit of about 100\AA\ in equivalent width for these
lines, which is larger than that for H$\beta$ line in the composite
quasar spectrum (46\AA; Vanden Berk et al. 2001).

\section{Methods of calculating SMBH masses}

\subsection{Using the broad-line width to calculate the SMBH viral
masses}

Assuming that the gas in BLRs is virial in the gravitational field
of the central SMBH, we can calculate the SMBH masses by the
following formula (e.g., Kaspi et al.2000, Kaspi et al. 2005):
\begin{eqnarray}
M_{\rm BH}=\frac{R_{\rm BLR}v_{\rm BLR}^{2}}{G}
\end{eqnarray}
where $G$ is the gravitational constant; $R_{\rm BLR}$ is the
distance from BLRs clouds to the central black hole, which can be
calculated from the monochromatic luminosity at $5100\AA\ (\lv)$
by the empirical size-luminosity relation, that is (Kaspi et al.
2005):
\begin{eqnarray}
R^{\rm \lv}_{\rm BLR}=(22.3\pm 2.1)\left(\frac{\lv}
{10^{44}~\ergs}\right)^{0.69\pm0.05}
\end{eqnarray}
where $v_{\rm BLR}$ is the BLRs virial velocity, which can be
traced by using FWHM of the broad $\hb$ line ($\rm FWHM_{\hb}$) or
$H_{\gamma}$ line, that is:
\begin{eqnarray}
v_{\rm BLR}=\sqrt{f}\times \rm FWHM_{\hb}
\end{eqnarray}
$f$ is the scaling factor. If BLRs cloud is disk-like with a
inclination of $\theta$ (Wills $\&$ Browne 1986; Collin et al.
2006):
\begin{equation}
{\rm FWHM_{\rm H \beta}}=2(v_{\rm r}^{2}+v_{\rm BLR}^{2} \rm
sin^{2}\theta)^{1/2}
\end{equation}
where $v_{r}$ is the random isotropic component. If assuming $v_{
r}\ll v_{\rm BLR}$ and the random orbits of clouds in BLRs, then
$f=0.75$, i.e., $\rm sin^2\theta=1/3$. Therefore, the formula (1)
can be transformed into the following form:
\begin{eqnarray}
M_{\rm BH} =0.75\times 4.35 \times 10^6 \left(\frac{\rm
 FWHM_{\hb}}{10^{3} \kms}\right)^2\left(\frac{\lv}{10^{44}\ergs}
\right)^{0.69}  \msun.
\end{eqnarray}

For type II AGNs, we could not obtain the intrinsic monochromatic
luminosity at 5100$\AA$ from their spectrum because of torus
obscuration. The \oiii$\lambda$5007 luminosity, $L_{\rm [O III]}$,
coming from NLRs is not obscured and is isotropic. Kauffmann et al.
(2003) have shown that the extinction-corrected $L^{cor}_{\rm [O
III]}$ is a good indicator of AGN activity for type II AGNs. The
unobscured monochromatic luminosity at 5100\AA\ is calculated by the
following relation (e.g., Kaspi et al. 2000, Heckman et al. 2004,
Netzer \& Trakhtenbrot 2007):
\begin{equation}
L_{\rm bol}=9 [\lv]=3500L^{cor}_{[\rm O III]}
\end{equation}

\subsection{ $\mbh- \sigma_*$ relation}

The second method for us to calculate the black hole masses is the
relationship with to the stellar velocity dispersion ($\sigma_*$)
in their host galaxies (Tremaine et al. 2002):
\begin{equation}
M_{\rm BH}(\sigma_{\ast})=10^{8.13}(\frac{\sigma_{\ast}}{200\kms})^{4.02}
\end{equation}

For 8 type II AGNs with redsshifts z$<$0.05, we obtained their
stellar velocity dispersions from the literature (see Table 1).
Other 4 type II AGNs in the redshift range 0.3$<$z$<$0.5 are
selected from the spectroscopic database of SDSS, their host
galaxies light is too weak to directly measure the stellar velocity
dispersion, $\sigma_*$. Because NLRs are primarily controlled by the
host gravitational potential, FWHM of core/narrow \oiii$\lambda$5007
line after removing asymmetric blue wing can be used to trace the
stellar velocity dispersion for these 4 type II AGNs (Nelson \&
Whittle 1996; Greene \& Ho 2005):
\begin{equation}
\sigma_*=\rm FWHM^{\rm core}_{\rm [O III]}/2.35
\end{equation}

\section{Data analysis}
For 4 type II AGNs with redshifts $0.3<z<0.5$, we downloaded their
spectrum from the spectroscopic database of SDSS and used following
steps to do the SDSS spectral measurements, which has been used for
SDSS type I AGNs. (1)First, we do the Galactic extinction in
observed spectra by using the extinction law of Cardelli, Clayton \&
Mathis (1989)(IR band) and O'Donnell (1994) (optical band),then the
spectra are transformed into the rest frame defined by the redshifts
given in their FITS headers. (2)The $\feii$template is from NLS1 I
ZW1 model, the Balmer continuum are calculated following
Grandi(1982). The best subtraction of the $\feii$, power-low and
Balmer continuum is found when $\chi^{2}$minimized in the fitting
windows: 3550-3645, 4170-4260, 4430-4770, 5080-5550, 6050-6220,
6890-7010$\AA$. Since they are the spectra of type II AGNs, the
$\feii$ line and Balmer continuum could be neglected, the power-law
continuum seems flat (see a sample fit in the a panel of Figure 1).
(3)Two sets of two-Gaussian are used to model \oiii$\lambda \lambda
4959, 6007$ lines. We take the same linewidth for each component,
and fix the flux ratio of \oiii$\lambda$4959 to \oiii $\lambda$5007
to be 1:3. Two-Gaussian profile is used to model \hb line. (see the
b panel of Figure 1).

From above spectral measurement, we obtain the FWHM of the
narrow/core \oiii\ line , as well as the total \oiii\ luminosity
(see table 1).

\begin{tiny}
\begin{table*}
\begin{center}
\begin{tabular}{lccccccccc}\hline \hline
name & z & log($L_{\rm [O III]}$)&$FWHM_{\rm [O III]}$&$\sigma_{\ast}$& $FWHM_{\hb}$&log($M_{\hb}/\msun$)&log($M_{\sigma_*}/\msun$)& Ref\\
     &   &  $\ergs$  &        $\kms$            & $\kms$      & $\kms$         &                     &                   &\\
(1) & (2) &(3) & (4) & (5) & (6) & (7) & (8)& (9)\\
 \hline
NGC1068 & 0.004 & 42.645$^a$ & -  & 128 & 3030 & 7.88 & 7.35 & $1,1,2$\\
NGC7212 & 0.027 & 42.636$^a$ & -  & 137 & 5420 & 8.37 & 7.47 & $1,1,2$\\
NGC7674 & 0.029 & 42.495$^a$ & -  & 144 & 2830 & 7.82 & 7.56 & $1,1,2$\\
Mrk3    & 0.014 & 43.221$^a$ & -  & 269 & 6000 & 9.17 & 8.65 & $1,1,2$\\
Mrk348  & 0.015 & 41.912$^a$ & -  & 185 & 9350 & 8.17 & 7.99 & $1,1,2$\\
Mrk463E & 0.050 & 42.785$^a$ & -  & 137 & 2770 & 7.81 & 7.88 & $1,1,2$\\
Mrk477  & 0.038 & 43.543$^a$ & -  & 117 & 4130 & 8.84 & 7.19 & $1,1,2$\\
Mrk1210 & 0.013 & 42.195$^a$ & -  & 77  & 3080 & 7.90 & 6.46 & $1,1,2$\\
J0842+3625 & 0.561 & 43.601$^b$ & 303.76 &129.30$^c$ & 4900($H_{\gamma}$) &8.99 &7.37 & $3,3,4$\\
J1039+6430 & 0.402 & 42.901$^b$ & 385.75 &164.15$^c$ & 3070 & 7.89 & 7.79 & $3,3,4$ \\
J1543+4935 & 0.512 & 42.789$^b$ & 709.72 &302.01$^c$ & 4300 & 8.19 &8.84 & $3,3,4$\\
J1641+3858 & 0.596 & 43.474$^b$ & 395.72 &168.39$^c$ & 3200($H_{\gamma}$) & 8.62 &7.83 &$3,3,4$\\
\hline
\end{tabular}
\caption{The properties for 12 type II AGNs with HBLRs. Col.(1),
name; Col. (2), redshift; Col.3, the \oiii luminosity; Col. (4),
FWHM of the narrow \oiii line; Col. (5), stellar velocity
dispersion; Col. (6), FWHM of H$\beta$ from HBLRs; Col.(7), SMBH
masses derived from the polarimetric broad line; Col.(8), SMBH
masses derived from the $\mbh - \sigma_*$ relation. Col.(9),
reference for Col.(3,5,6):(1)Bian \& Gu 2007; (2) Tran 1995; (3)
this work; (4) Zakamska et al. 2005. Note: $^a$: the intrinsic
extinction-corrected \oiii luminosity;$^b$: the intrinsic
extinction-uncorrected \oiii luminosity; $^c$: the stellar
velocity dispersion is derived from the FWHM of the narrow/core
\oiii.}

\end{center}
\end{table*}
\end{tiny}

\begin{figure}
\begin{center}
\includegraphics[height=7cm,angle=-90]{f1a.eps}
\includegraphics[height=7cm,angle=-90]{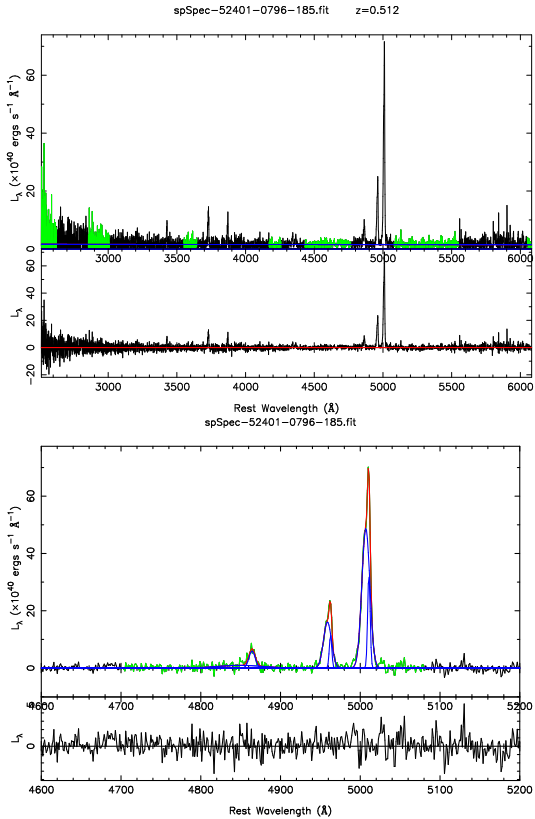}
\caption{Sample of SDSS spectrum measurement for J 1543+4395. In the
top panel, the black curve is the observed spectrum, the green
ranges are our fitting windows. In the bottom panel, the red line is
the sum of all multi-Gaussians (blue curves),  the green curve is
our fitting range of the pure \hb and \oiii emissions after the
subtraction of the power-law continuum, the Balmer continuum and Fe
multiples.}
\end{center}
\end{figure}

\section{Results and Discussion}
\subsection{Results}

In Table 1 we presented our results. We used formulas (5) and (6)
to calculate the virial SMBHs masses, which is listed in Col. (7)
in Table 1. By the $\mbh- \sigma_*$ relation, we used formulas (7)
and (8) to calculate the SMBHs masses from $\sigma_*$, which is
listed in Col. (8) in Table 1. We find that, except two objects,
Mrk463E and SDSS J1543+4395 (see Figure 2), the black hole masses
from the broad line width in spectro-polarimetric observations are
larger than that from the $\mbh- \sigma_*$ relation. The mass from
the polarimetric broad line width is averagely larger than that
from $\mbh- \sigma_*$ relation by about 0.6 dex.

For SDSS J1543+4935, the mass from the $\mbh- \sigma_*$ relation
is larger that from the polarimetric broad line width by 0.65 dex
(see Table 1). We note that, respect to other three objects, it is
difficult to define its narrow/core component from the \oiii
$\lambda5007$ emission line profile (see Figure 1 and Figure 3).
Here we used the FWHM of its component with larger flux as the
tracer of the $\sigma_*$. The dynamics of NLRs clouds for SDSS
J1543+4935 is probably special. If neglecting SDSS J1543+4395, we
find the the average mass difference would be larger, from 0.6 dex
to 0.7 dex.

\begin{figure}
\begin{center}
\includegraphics[height=6cm]{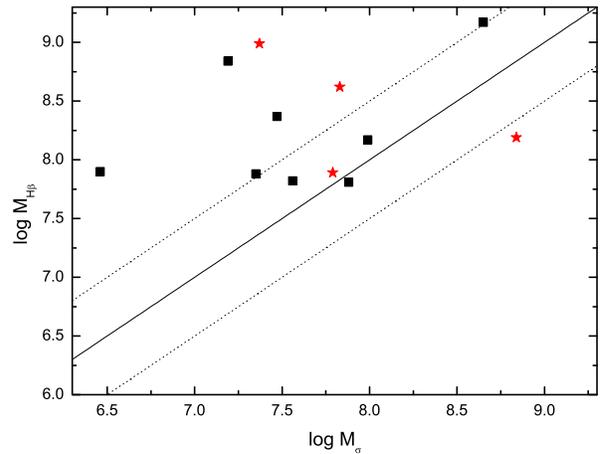}
\caption{log$\mbh$ vs log$M_{\sigma_*}$. $\mbh$ is from the
polarimetric broad-line width, $M_{\sigma_*}$ is calculated from
the $\mbh- \sigma_*$. Black square denotes the type II AGNs with
redshifts z$<$0.05, red star denotes the type AGNs with
0.3$<$z$<$0.8. The solid line denotes 1:1 and the dashed lines are
$y=x\pm 0.5$.}
\end{center}
\end{figure}

\begin{figure}
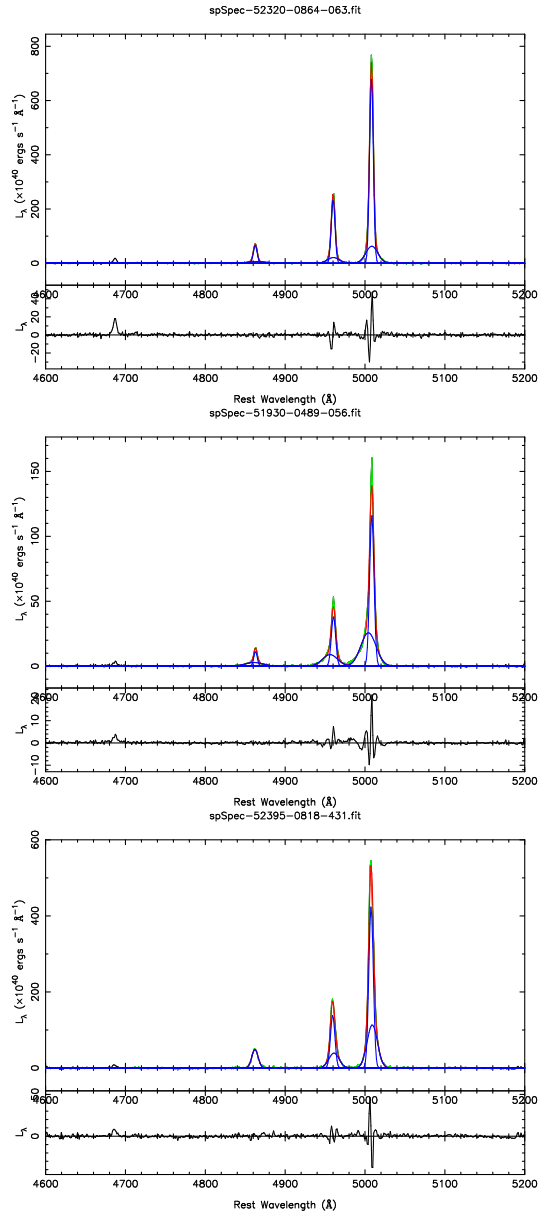

\begin{center}
\includegraphics[height=7cm,angle=-90]{f3a.eps}
\includegraphics[height=7cm,angle=-90]{f3b.eps}
\includegraphics[height=7cm,angle=-90]{f3c.eps}
\caption{The \oiii\ double line fitting for SDSS J0842+3625, SDSS
J1039+6430 and SDSS J1641+3858 (from top to bottom).}
\end{center}
\end{figure}

\subsection{Discussion}

The uncertainties in the black hole masses estimated by the $\mbh-
\sigma_*$ relation depend upon the uncertainties in the stellar
dispersion and the $\mbh- \sigma_*$ relation. For the typical
uncertainty of 20 \kms \ for $\sigma_{*}=200$ \kms, the error of
$\log~ \sigma_{*}$ would be about 0.05 dex, corresponding to 0.2
dex for $\log M_{\rm bh}$. Considering the error of 0.3 dex form
the $M_{\rm bh}-\sigma_{*}$ relation (Tremaine et al. 2002), the
error of $\log M_{\rm bh}$ is about 0.4 .

For the mass estimated by the polarimetric broad-line width, the
accuracy of the results depend on the FWHM measurement of the
broad $\hb$, the unobscured 5100\AA luminosity, the BLRS dynamics
and the empirical size-luminosity relation. The usage of
size-luminosity relation of Bentz et al. (2006) is almost the same
to that of Kaspi et al. (2006) for mass larger than $10^7$ \msun.
It is generally believed that the uncertainty from this method is
about 0.5 dex (e.g., Bian \& Gu 2007). The spectral resolution in
the polarimetric observation is about 8\AA\ for objects in Tran
(1995) and 19\AA\ for objects in Zakamaska et al. (2005). For our
sample, we find that the instrumental correction in FWHM will lead
to the mass uncertainty less than 0.1 dex.

For the 12 type II AGNs with polarimetric broad emission lines, we
find that the $\mbh$ derived from the broad-line width is generally
larger than that from the $\mbh-\sigma_*$ relation, with an average
of 0.6 dex higher. The intrinsic extinction-uncorrected \oiii
luminosity for 4 SDSS objects is the lower limit. Considering the
intrinsic correction, the mass derived from polarimetric broad-line
width would be larger. Because the mass from the broad-line width is
$\propto \rm FWHM_{\rm H\beta}^2~\lv^{0.69}$, the larger mass
derived from polarimetric broad-line width is mainly due to the
overestimate of virial velocity of SMBHs estimated from \hb FWHM
versus formula (4). It is possible that the profile from HBLR is
broadened in the polarimetric observation. There exists other
nonvirial broaden effect for the polarimetric broad line. If it is
not the case, considering the formula (4), $\rm FWHM_{\rm H
\beta}=2(v_{\rm r}^{2}+v_{\rm BLR}^{2} \rm sin^{2}\theta)^{1/2}$, it
is possible that the random isotropic velocity can't be neglected.
Therefore, there exists strong random isotropic velocity in BLRs,
such as outflow in HBLRs. The speeds of outflow expanding away from
the accretion disk may be comparable to the virial velocity of the
clouds in HBLR (Yong et al. 2007), the profile from HBLR may be
broaden due to the scatter in the outflow. Considering the random
orbits of BLRs clouds, $\rm sin^2\theta=1/3$, and $v_{r} \sim v_{\rm
BLR}$, the mass from the polarimetric broad lines FWHM would
decreased by 0.6 dex, which will make the mass from polarimetric
broad lines FWHM consistent with that from the $\mbh-\sigma_*$
relation.

For type II AGNs, the central engine and BLRs are not viewed
directly due to obscuration by the circum-nuclear torus. They are
seen at a large torus inclination to the line of sight. We are not
sure whether BLRs are coplanar with the torus. If they are coplanar,
type II AGNs are seen at a large BLRs inclination to the line of
sight. By $\rm sin^2\theta = 1$, i.e., $"edge-on"$, the mass from
equation (5) would decreased by 0.48 dex respect to that for $\rm
sin^2\theta=1/3$, which is still averagely larger than that from
$\mbh-\sigma_*$ relation. When the random isotropic velocity can't
be neglected, e.g., $v_{r} \sim v_{\rm BLR}$, the $"edge-on"$ BLRs
geometry would lead the mass from equation (5) decreased by 0.18 dex
respect to the random BLRs clouds. In this case, the inclination is
not the dominant source of the larger SMBH masses deviation between
the mass from the width of polarimetric broad emission lines and
that from the $\mbh-\sigma_*$ relation, while the large random
velocity is the dominant source (see also Collin et al. 2006).

\section{Conclusion}
The central SMBHs masses for 12 type II AGNs with polarimetric broad
lines are estimated from the broad line FWHM and the \oiii
luminosity, as well as the $\mbh-\sigma_*$ relation. We find that:
(1)the SMBH masses derived from the width of polarimetric broad
emission lines is averagely larger than that from the
$\mbh-\sigma_*$ relation by about 0.6 dex, (2)If these type II AGNs
follow $\mbh-\sigma_*$ relation, our result suggested that the
random velocity can't not be omitted and is comparable with the BLRs
Keplerian velocity. It is consistent with the scenery of larger
outflow from the accretion disk suggested by Yong et al. (2007).

{\noindent \rm {Acknowledgements}. We thank the anonymous referee
for helpful suggestions. This work has been supported by the NSFC
(No. 10733010), the Science-Technology Key Foundation from Education
Department of P. R. China (No. 206053), and the China Postdoctoral
Science Foundation (No. 20060400502).}

\end{document}